\documentclass{icrc2009}

\usepackage{graphicx}   
\usepackage[caption=false]{caption}    
\usepackage[font=footnotesize]{subfig} 
\usepackage{fixltx2e}
\usepackage{url}

\newcommand{\shorttitle}[1]%
{\markboth{Proceedings of the 31\MakeLowercase{$^{st}$} ICRC, {\L}\'{o}d\'{z} 2009}{#1} }
\newcommand{\etal}{\MakeLowercase{\textit{et al. }}} 

\usepackage{amssymb}
\def\bd{\begin{displaymath}}
\def\be{\begin{equation}}
\def\ed{\end{displaymath}}
\def\ee{\end{equation}}

\begin{document}
\title{Analysis of the Spectral Intensities and Ratios of Electrons and Positrons in Cosmic Rays}

\author{\IEEEauthorblockN{R. Cowsik and B. Burch}
                            \\
\IEEEauthorblockA{McDonnell Center for the Space Sciences and Physics Department}
\IEEEauthorblockA{Washington University, St. Louis, MO 63130}}

\shorttitle{R. Cowsik \etal Cosmic Ray Electrons}
\maketitle

\begin{abstract}
The observations of the total electronic component and the positron fraction in cosmic rays by the FERMI, HESS, ATIC, and PAMELA instruments are studied with analytical propagation models, both for a set of discrete sources and for a spatially smooth source distribution. The positron fraction over the entire energy range of $\sim1-100~GeV$ is shown to fit with the nested leaky box model. We derive the spectrum of electrons in cosmic rays arising from direct acceleration by the sources and discuss the narrow spectral feature in the spectrum.
  \end{abstract}

 
\section{Introduction}
\begin{figure}
  \centering
  \includegraphics[width=2.5in]{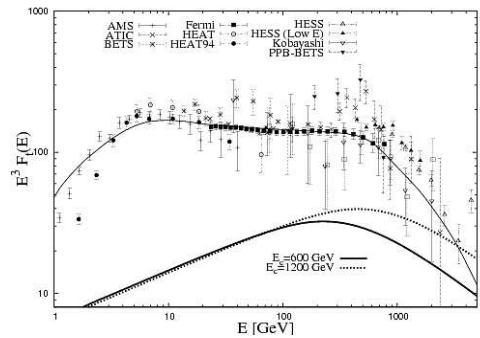}
  \caption{The measurements of the total electron spectrum, a smooth fit (solid line), and the total secondary electron and positron spectrum calculated from the nested leaky box model (thick solid and dotted lines) for $E_c=600~GeV$ and $E_c=1200~GeV$.}
  \label{HESS}
 \end{figure}

\begin{figure}
  \centering
  \includegraphics[width=2.5in]{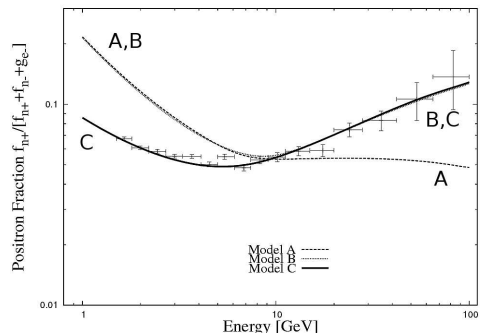}
  \caption{The theoretically calculated positron fraction in models A (similar to that of Moskalenko and Strong \cite{Moskalenko}), B, and C are compared with the observations. All calculations are normalized at $\sim10$ GeV.}
 \label{PAM}
\end{figure}

Progressive improvements of detectors since the pioneering efforts in the 1960s have led to improved measurements of the spectrum of the electronic component and positron fraction by the FERMI, HESS, ATIC, and PAMELA instruments [1-5]. These measurements have stimulated great excitement, not only because of their importance to cosmic ray studies, but also because of the possible interpretation of some of the features from signals from the annihilation of dark matter in the Galaxy. We reproduce the observed total electronic spectrum $f_t(E)$ and the positron fraction $R(E)$ in fig. \ref{HESS} and fig. \ref{PAM} respectively.

In section II, we show that these observations help us in distinguishing between the leaky box \cite{Cowsik67} and the nested leaky box models \cite{Cowsik75}. It is the difference in the kinematics of the reactions producing secondary `light' nuclei and positrons that helps in distinguishing amongst these models. We show that the positron fraction observed by PAMELA may be fit well with the nested leaky box model. We begin section III by deriving the spectrum of electrons resulting from direct acceleration in the cosmic ray sources by subtracting the secondary electrons and positrons from the spectrum of the total electronic component. After this, the  rest of the section is devoted to a discussion of this primary electron spectrum in terms of a set of discrete sources of cosmic rays sprinkled throughout the Galaxy. The analysis indicates that we need the closest source to be no more than $\sim$ 200 pc from the solar system, which in turn implies that there are about 5000 sources accelerating cosmic rays in the Galaxy. The contrast between the spectra expected from a discrete set of sources and from sources spatially distributed smoothly throughout the Galaxy plays an important role in analyzing signatures from the annihilation of dark matter in the Galaxy. This aspect is discussed in section IV. It would be appropriate to note here that the theoretical considerations of cosmic ray transport with exponential pathlength distributions \cite{Cowsik67}\cite{Cowsik66} and its analysis in terms of a set of discrete sources sprinkled over the galaxy \cite{Cowsik75}\cite{Cowsik73} started three or four decades ago. The recent improved measurements constitute a testing ground for these early ideas on cosmic ray physics. 


\section{Positron Fraction in Cosmic Rays and the Leaky Box Models}
\begin{figure}
  \centering
  \includegraphics[width=2.5in]{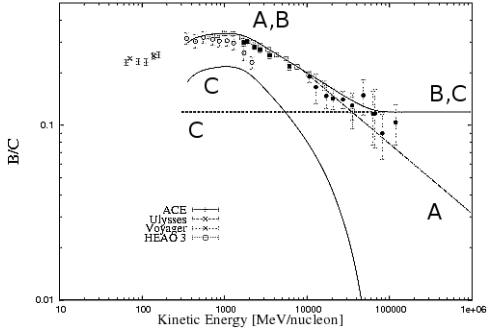}
  \caption{The observed B/C secondary to primary ratio is plotted (points from a compilation in \cite{Strong}) along with the power law extrapolation at high energies (dot-dashed line, Model A), a constant extrapolation (solid line, Model B), and a two-component fit (dotted lines, Model C).}
  \label{BC}
 \end{figure}
The observations of decreasing ratios of secondaries to primaries like B/C and (V+Ti+Sc)/Fe with increasing energy per nucleon in cosmic rays could be incorporated into the original leaky box model (A and B) \cite{Cowsik67} by letting the mean residence time $\tau$ depend on energy in such a way as to reproduce the observations \cite{Strong}. A viable alternative to this idea was the nested leaky box model (C) \cite{Cowsik75} in which the energy dependant part of the ratio (proportional to $\tau_s(E)$) was attributed to the spallation in the matter constrained in a cocoon-like region surrounding the sources and the resident time in the Galaxy $\tau_G$ was assumed independent of energy. With observations of the ratios limited to a finite range of energies, it became customary to extrapolate the residence times as a simple power law, decreasing as $\sim\tau E^{-\Delta}$. Even though the observations to date allow $\Delta\sim0$ beyond some tens of GeV/n, this alternative had not been discussed earlier. These three possibilities and their corresponding fit to the secondary to primary ratios are shown in fig. \ref{BC} for the B/C ratio. The (V+Ti+Sc)/Fe ratio can be fit similarly. The three models presented here are characterized by mean residence times as noted below:
\be \textrm{A:}\left.\begin{array}{ll}
\tau_A(E)&\sim\tau_0E^{-\Delta},~~~~~~~E>2~GeV/n~~~~
\end{array}\right.\ee
\be\textrm{B:}\left.\begin{array}{ll}
\tau_B(E) & \sim\tau_A(E),~~~~~~~~~E\lesssim10~GeV/n\\~~
& \sim\tau_G\sim X,~~~~~~~E\gtrsim10~GeV/n~~\end{array}\right.\ee
\be \textrm{C:}\left.\begin{array}{ll}
\tau_s(E)&\sim\tau_B(E)-\tau_G,~~~E\lesssim10~GeV/n\\
&\sim\tau_G\sim X,~~~~~~~~E\gtrsim10~GeV/n\end{array}\right.\ee
where $X$ is a constant.

Consider now the injection of cosmic rays by sources with the spectral form \be Q(E)\sim Q_0E^{-\beta}cm^{-3}s^{-1}sr^{-1}GeV/n^{-1}.\ee When spallation and energy loss during transport could be neglected, the three models led to the following equilibrium spectra of cosmic rays in the Galaxy:
\be\label{eq:f}\begin{array}{ll}
 \textrm{A:}~f_A&\sim Q_0\tau_0E^{-(\beta+\Delta)}\\
 \textrm{B:}~f_B&\sim Q_0\tau_0E^{-(\beta+\Delta)},~~~~E\lesssim10~GeV~\nonumber\\
 &\sim Q_0\tau_GE^{-\beta},~~~~~~~~~E\gtrsim10~GeV\\
 \textrm{C:}~f_C&\sim Q_0\tau_GE^{-\beta}. \end{array}\ee
Note that $f_C$ does not depend upon $\tau_s(E)$ \cite{Cowsik75}.

Unless $\beta$ itself is a function of energy so that eq. 5B reproduces the observed spectra of the nuclear component that is a simple power law, model B is not viable. Model A requires $\beta\sim2.22$ and predicts increasing anisotropy with energy of cosmic rays, and model C requires $\beta\sim2.65$ and generates constant anisotropy at all energies. 

The observations of the positron fraction provides a clear way of choosing amongst these three models. The reason for this is related to the fact that in the spallation process that generates the secondary cosmic ray nuclei like B in collisions of the primary C nuclei, the daughter nuclei will emerge with the same energy per nucleon as the parents. On the other hand, the production of positrons proceeds through the production of mesons (mainly pions) in the collision of the primary nuclei which follow the decay chain $\pi^\pm\rightarrow\mu^\pm+\nu_\mu(\bar{\nu}_\mu)$, followed by $\mu^\pm\rightarrow e^\pm+\nu_e(\bar{\nu}_e)+\bar{\nu}_\mu(\nu_\mu)$. In this process, the positrons and the secondary electrons carry, on the average, only a small fraction $\sim0.05$ of the energy/nucleon of the primary. As a consequence of this, the production spectrum of positrons (and secondary electrons) in all the three models are nearly identical. The fact that low energy cosmic rays spend more time in the cocoon surrounding the sources plays no role as the $e^+$ and $e^-$ secondaries are generated only by the high energy part of the nucleon spectrum. Now the equilibrium spectrum of secondary $e^+$ and $e^-$ predicted by the three models simply follow the product of the production rate, $Q_{e\pm}(E)$ and the residence time $\tau(E)$ at low energies. At very high energies, all the models predict steeper spectra $\sim Q_{e\pm}(E)E^{-1}$:
\begin{eqnarray}\label{eq:pos}
\textrm{A:}~f_{n\pm}(E)&=&Q_{n\pm}(E)\tau_A(E)\nonumber\\
&\sim&\tau_0E^{-(\beta+\Delta)},~~~~~~~~~E\ll E_c\nonumber\\
&\sim&E^{-(\beta+1)},~~~~~~~~~~~~E\gg E_c\nonumber\\
\textrm{B:}~f_{n\pm}(E)&=&Q_{n\pm}(E)\tau_B(E)\nonumber\\
&\sim&E^{-(\beta+\Delta)},~~~~~~E\lesssim10~GeV\nonumber\\
&\sim&E^{-\beta},~~~~10~GeV\lesssim E\ll E_c\nonumber\\
&\sim&E^{-\beta+1},~~~~~~~~~~~~~~E\gg E_c\nonumber\\
\textrm{C:}~f_{n\pm}(E)&=&Q_{n\pm}(E)\tau_G(E)\nonumber\\
&\sim& E^{-\beta},~~~~~~~~~~~~~~~~~E\ll E_c\nonumber\\
&\sim& E^{-\beta+1},~~~~~~~~~~~~~~E\gg E_c
\end{eqnarray}
where $E_c\approx1/b\tau_G$ and $b$ is defined in eq. 10. In fig. \ref{PAM} we show the positron fraction 
\be R(E)=f_{n+}(E)/f_t(E)\ee
with the positron spectrum $f_{n+}(E)$ as given in eq. 6 for the three models, and the spectrum of the total electronic component is taken to be the smooth fit to the data shown in fig. \ref{HESS}. Note that only model C provides a good fit over the entire range of the PAMELA observations. Thus, the observed positron fraction resolves the degeneracy amongst models developed to understand the nuclear secondaries and suggests that the nested leaky box model may provide a closer approximation to cosmic ray transport in the Galaxy \cite{Cowsik09}.

\section{Spectrum of Electrons Generated by Cosmic Ray Sources}
\begin{figure}
  \centering
  \includegraphics[width=2.5in]{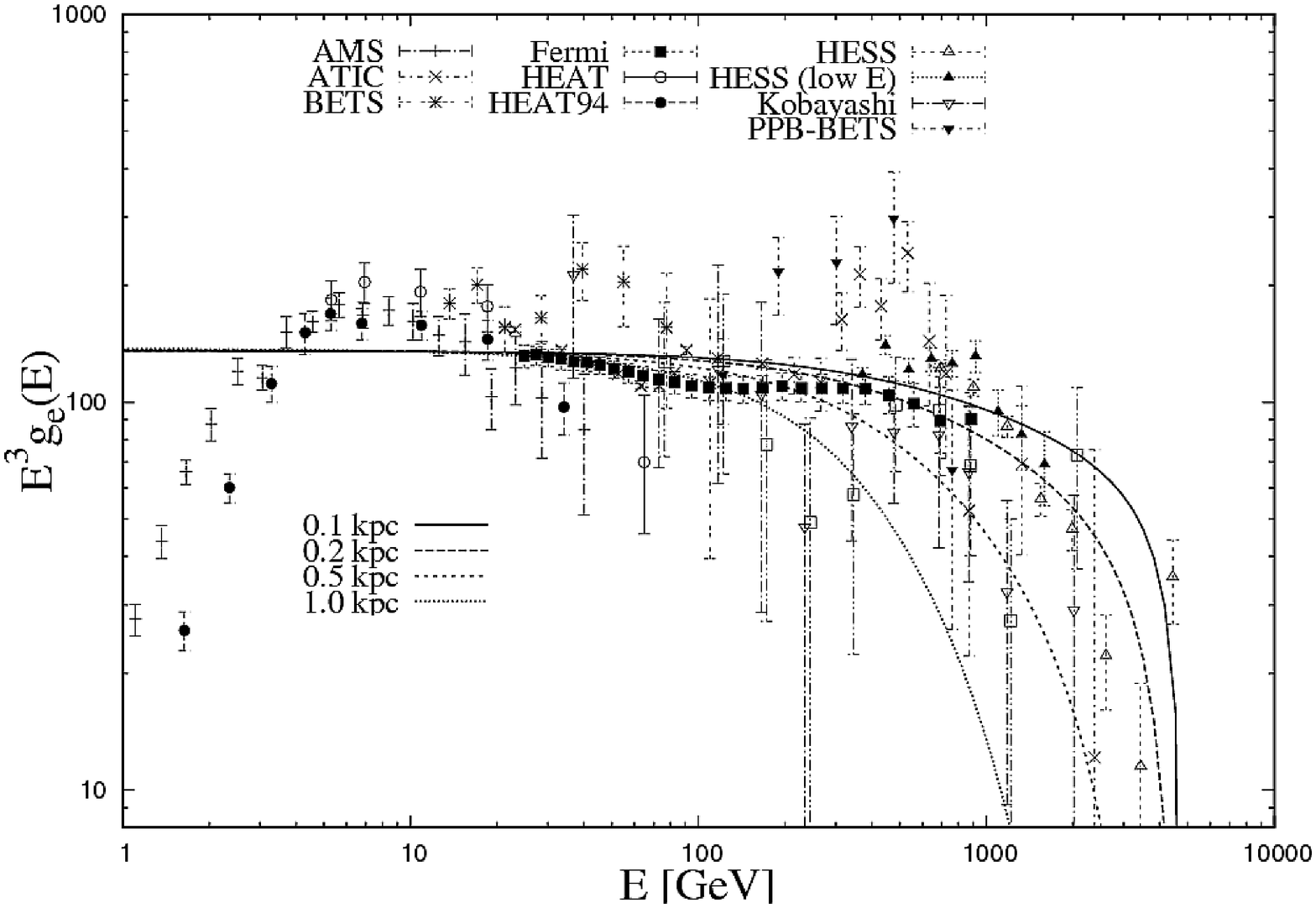}
  \caption{The primary electron spectra due to a single source at various distances from the source with $E_x=5~TeV$ compared to the \emph{primary} electron spectrum.}
  \label{SINGLE}
\end{figure}

\begin{figure}
  \centering
  \includegraphics[width=2.5in]{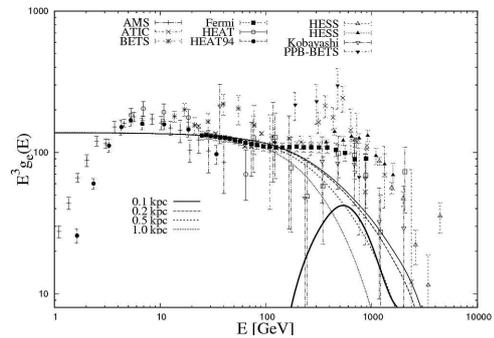}
  \caption{The primary electron spectra resulting from many cosmic ray sources at various distances to the first source with $E_x=5~TeV$ compared with the \emph{primary} electron spectrum. Their difference (thick line) is shown.}
  \label{SUM}
 \end{figure}

The nucleonic component in the primary cosmic rays consists mainly of protons with a small fraction of neutrons which come in bound to He and other nuclei. As a consequence, the production of positrons is favored over that of electrons in cosmic ray secondaries:
\be \eta\equiv Q_{n-}(E)/Q_{n+}(E)\approx0.5-0.8.\ee
The lower value of $\eta$ is favored by the theoretical estimates based on the data on nuclear interactions obtained at accelerators \cite{Protheroe}, and the larger value is favored by the observations of the $\mu^-/\mu^+$ ratio generated by cosmic rays in the atmosphere \cite{Hayakawa}. The loss of energy by energetic cosmic ray electrons due to inverse Compton scattering and synchrotron emission in the magnetic fields prevalent in the propagation volume is given by 
\be dE/dt=-bE^2;\ee
\begin{eqnarray} 
b&=&\Big(3.22\times10^{-3}\Big(\frac{\omega_{ph}}{eV\cdot cm^{-3}}\Big)\nonumber\\
&&+7.9\times10^{-2}H^2_{\mu Gauss}\Big)GeV^{-1}Myr^{-1}.
\end{eqnarray}
Noting that the scattering of starlight by the very high energy electrons is described by the Klein-Nishina formula, the effect of starlight may be neglected, and only the microwave background contributes to $\omega_{ph}$ in eq. 10. Taking $H\sim5~\mu Gauss$ and $\omega_{ph}\approx0.25eV~cm^{-3}$, we get $b\approx1.56~GeV^{-1}Myr^{-1}$. Adopting the nested leaky box model, the spectrum of the secondary positrons and electrons is given by \cite{Cowsik66}
\begin{eqnarray} 
f_{n+}(E)&=& f_{n-}(E)/\eta\\
&=&\int_0^{1/bE}Q_0E^{-\beta}(1-bEt)^{\beta-2}e^{-t/\tau_G}dt.\nonumber\end{eqnarray}
In fig. \ref{HESS}, we show the total secondary component $f_s(E)=f_{n+}(E)(1+\eta)$, with the normalization for $Q_0$ determined from a smooth fit to the PAMELA data and $\tau_G\sim1Myr$, i.e. $E_c=1/b\tau_G\approx600~GeV$. We also show the secondary component for $E_c\approx1200~GeV$. Subtracting this (for $E_c=600Gev$) from the total spectrum of the electronic component, we get the spectrum due to the sources, 
\be g_e(E)=f_t(E)-f_{n+}(E)(1+\eta),\ee
which is displayed in figs. \ref{SINGLE} and \ref{SUM}.

In an attempt to understand this spectrum, we assume diffusive transport of cosmic rays and write the differential equation describing the transport as
\be dN/dt-\kappa\nabla^2N+N/\tau=Q.\ee
Note that here we have simplified the form of the Green's function by introducing an escape term $t/\tau$ in place of the boundary condition demanding that the cosmic ray density vanish at the planar surfaces of the thick cosmic ray disk of the Galaxy. The Green's function is given by
\be G(r,t)=(4\pi\kappa t)^{-3/2}exp\Big(-\frac{r^2}{4\kappa t}-\frac{t}{\tau}\Big).\ee
This along with the subsidiary eq. 9 is adequate to describe the transport of the electronic component of cosmic rays. Note that for a smooth and uniform distribution of sources $Q(r,t)=\delta(t)$, the Green's function in eq. 14 integrates to the simple leaky box model. 


Suppose we have a source situated at a distance $r_i$ which is continuously emitting cosmic ray electrons with a spectrum
\be\begin{array}{ll}
Q&\sim~Q_0E^{-\Gamma},~~~~~~E<E_x\\
&\sim~0,~~~~~~~~~~~~~~E\geq E_x,
\end{array}\ee
the observed spectrum is given by
\be\begin{array}{ll}
f_d(E,r_i)=&\\
~~~~~\int^{\frac{E_x-E}{bEE_x}}_0Q_0E^{-\Gamma}(1-bEt)^{\Gamma-2}G(r_i,t)dt.&\end{array}\ee
We display in fig. \ref{SINGLE} this spectrum for various values of $r_i$. Note that for $\Gamma\approx3$, $E_x\approx5-10~TeV$, and $r_i\approx0.2-0.5~kpc$, one can obtain a reasonable fit to the high energy part of the primary electron spectrum g(E).

However, it is appropriate to add the contributions of all the sources situated at different distances to obtain their net contribution $f_D(E)$ to $g_e(E)$. Assuming that the $i^{th}$ source is at an average distance of $\sqrt{i}r_1$, we sum their contributions and compare with the primary electron spectrum $g_e(E)$ in fig. \ref{SUM}. We see that irrespective of the choice of parameters, the high energy part of $g_e(E)$ is not reproduced by the theoretical expectation $f_D(E)$. This difference becomes particularly large for $r_1$ much larger than $\sim200~pc$. Thus we may expect that the distance to the nearest source and the typical spacing between the sources is $\sim200~pc$. The difference between the contribution of the discrete sources $f_D(E)$ and the fit to $g_e(E)$ is also shown in fig. \ref{SUM}, and the difference $n_e(E)$ with the data points is shown in fig. \ref{CONTELOSS}.

\section{Narrow Spectral Features in the Primary Electron Spectrum}
\begin{figure}
  \centering
  \includegraphics[width=2.5in]{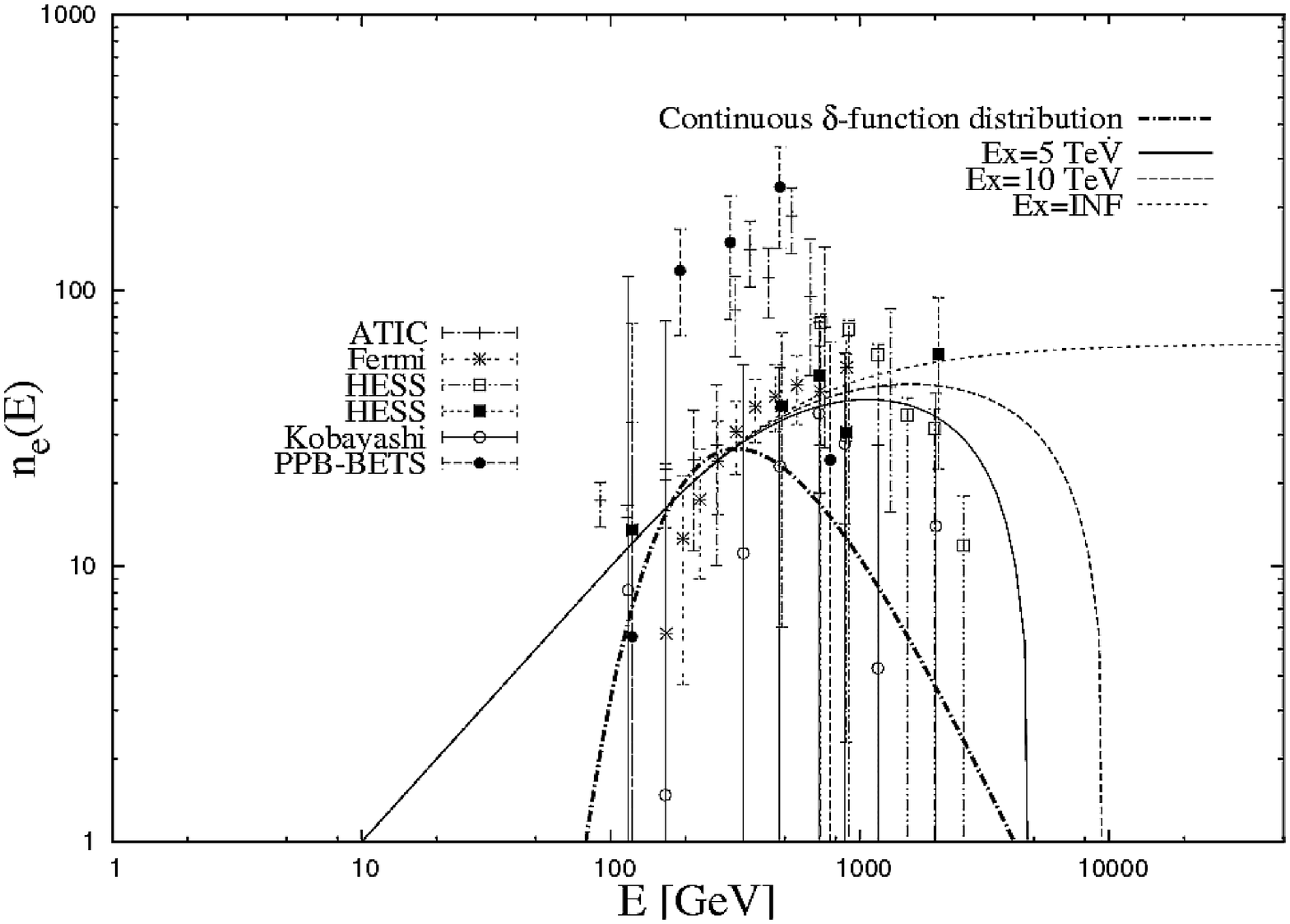}
  \caption{The spectrum of electrons accelerated in shocks to fit the electron excess for a cutoff energy between $5-10~TeV$ (thin lines). Also plotted is the spectrum for a $\delta$-function input from continuous distribution of sources with $E_a=1200~GeV$. Note the peak at $\sim300~GeV\approx E_c/2$ (thick dot-dashed line).}
  \label{CONTELOSS}
 \end{figure}

The narrow spectral feature displayed in figs. \ref{SUM} and \ref{CONTELOSS} has been ascribed \emph{in toto} or in part to products of dark matter annihilation (see \cite{Profumo} for references). In this section, we discuss
two input spectra, one a $\delta$-function in energy and the other a flat spectrum $\sim E^{-2}$, such as that expected for acceleration at planar shocks of high Mach number. 

\begin{figure}
  \centering
  \includegraphics[width=2.5in]{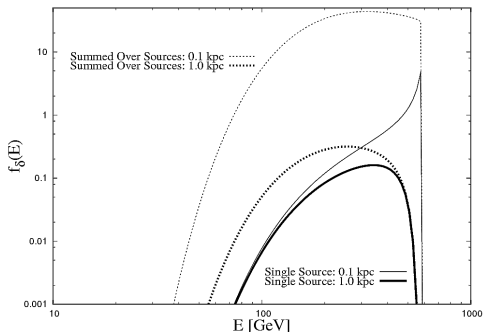}
  \caption{Shown here is the spectrum from a $\delta$-function input from a single source (solid lines) and the spectra from the sum over a discrete distribution of $\delta$-function sources (dotted lines) for various distances.}
  \label{DELTA}
 \end{figure}

\subsection{$\delta$-function Input Spectrum}
Here, the injection spectrum $Q(t=0,E)$, the spectrum after time $t$, $f_{\delta 1}(E)$ (without leakage), and $f_{\delta c}(E)$ for a spatially smooth source and leakage lifetime $\tau_G$ are given by
\be Q(t=0,E_a)=\delta(E(t=0)-E_a),\ee
\be f_{\delta 1}(E)=\frac{E_a^2}{E^2}\delta\Big(\frac{E}{1-bEt}-E_a\Big),\ee
\be f_{\delta c}=\frac{1}{bE^2}exp-\Big(\frac{E_a-E}{bE_aE\tau_G}\Big).\ee
The spectrum $f_{\delta c}(E)$ peaks at $E=E_c/2=1/b\tau_G$ for $E_a>E_c/2$, and there is no sharp peak at $E_a$ (see fig. \ref{CONTELOSS}). A $\delta$-function input from discrete sources leads to an observed spectrum
\bd f_{\delta d}(E,r_i)=\frac{E_a^2}{E^2}\Big(\frac{bE_aE}{4\pi\kappa(E_a-E)}\Big)^{3/2}~~~~~~~~~~~~~~~~~\nonumber\ed
\be~~~~~~~~~~~~~~~~~~~~~~~\times exp-\Big(\frac{br_i^2EE_a}{4\kappa(E_a-E)}+\frac{E_a-E}{bE_aE\tau_G}\Big).~\ee
For large $\tau_G$, this spectrum displays a peak at 
\be E_{peak}\approx\frac{6\kappa E_a}{6\kappa+br^2_iE_a}.\ee
For small $r_i$, the peak will be sharp near $E_a$, but with increasing $r_i$, the peak will shift to lower energies and will become broader. We display in fig. \ref{DELTA} some examples of the spectra generated by $\delta$-function inputs at various $r_i$ and show the sum, $f_D(E)$, over the sources at various distances in fig. \ref{SUM}.

\subsection{Shock Acceleration}
Planar shocks of high Mach number yield
\be Q_{shock}(E)\sim Q_0E^{-2},~~~~~~~~~E<E_x,\ee
resulting in an equilibrium spectrum
\be f_2(E)=\frac{\tau Q_0}{E^2}\bigg(1-e^{-\frac{E_x-E}{bEE_x\tau_G}}\bigg),\ee
which can reproduce $n_e(E)$ as shown in fig. \ref{CONTELOSS}.

We refer the reader to ref. \cite{Profumo} for a comprehensive overview of current efforts to explain these observations and to ref. \cite{Burch09} for more details regarding this paper.

\end{document}